\documentclass[12pt]{article} 
\usepackage{amssymb}
\usepackage{graphicx}
\begin{document} 
\begin{center}
{\large \bf Cherenkov and Fano effects at the origin of asymmetric vector
mesons in nuclear media}

\vspace{0.5cm}                   

{\bf I.M. Dremin}

\vspace{0.5cm}                       

         Lebedev Physical Institute, Moscow 119991, Russia\\
\medskip

        National Research Nuclear University "MEPhI", Moscow 115409, Russia     

\end{center}

\begin{abstract}

It is argued that the experimentally observed phenomenon of asymmetric vector 
mesons produced in nuclear media during high energy nucleus-nucleus collisions 
can be explained as Cherenkov and Fano effects. The mass distributions of 
lepton pairs created at meson decays decline from the traditional Breit-Wigner 
shape in the low-mass wing of the resonance. That is explained by the positive 
real part of the amplitude in this wing for classic Cherenkov treatment and 
further detalized in quantum mechanics as the interference of direct and 
continuum states in Fano effect. The corresponding parameters are found from 
the comparison with $\rho $-meson data and admit reasonable explanation.
\end{abstract}

Resonance peaks are observed in many natural phenomena. The traditional way
of their description is to compare their shapes with the symmetric
Breit-Wigner formula \cite{bw}
\begin{equation}
f=\frac {k}{(m_{r }^2-M^2)^2+m_r^2\Gamma ^2},
\label{bw}
\end{equation}
where $f$ denotes a signal strength, $k$ is the normalization constant, 
$M$ is the scanning energy, $m_r$ is the maximum position (the resonance
mass), $\Gamma $ is the resonance width. One measures the signal intensity
at different energies. For example, the atoms behaving as oscillators emit 
as Breit-Wigner resonances. The resonance shape is a purely statistical
phenomenon. It depends on many details of interactions and need not to be
necessarily symmetric.

The asymmetric resonance peaks were experimentally observed in various
fields of physics even before the formula (\ref{bw}) was proposed. 
E.g., the resonance of He atoms observed in the inelastic scattering of
electrons is strongly asymmetric (see \cite{rice, fano}). 

In particle physics, such peaks are identified with unstable particles. 
The symmetric shape is observed for resonances directly produced in particle
collisions. Their characteristics are compiled by the PDG (Particle Data 
Group). The situation has changed after the high energy nucleus-nucleus 
collisions became available. The created particles have to leak somehow from
the nuclear medium. Medium interactions may lead to some modification of their
characteristics. Really, there are numerous experimental data 
\cite{agak, adam, arna, damj, trnk, naru, muto, kozl, kotu, tser} about the
in-medium modification of widths and positions of prominent vector meson
resonances. Some of them even contradict each other.
They are mostly obtained from the shapes of dilepton 
(decay products) mass and transverse momentum spectra in nucleus-nucleus 
collisions. The dilepton mass spectra decrease approximately exponentially
with increase of masses but show peaks over this trend at some masses which
can be identified with prominent resonances. The $\rho $-meson peak is usually
the strongest one \cite{agak, adam, arna, damj} in the ratio 
$\rho :\omega :\phi =10:1:2$. Below, we concentrate on properties of
in-medium $\rho $-mesons.

The dilepton mass spectrum in semi-central In-In collisions at 158 AGeV
measured by NA60-Collaboration \cite{arna} is shown in Fig. 1
by dots with error bars in the region of $\rho $ and $\omega $-mesons. 
Its asymmetry is easily seen with some excess in the low-mass wing. The shape 
is quite distinct from the familiar $\rho $-peak.

\begin{figure}[h]
\includegraphics[width=\textwidth]{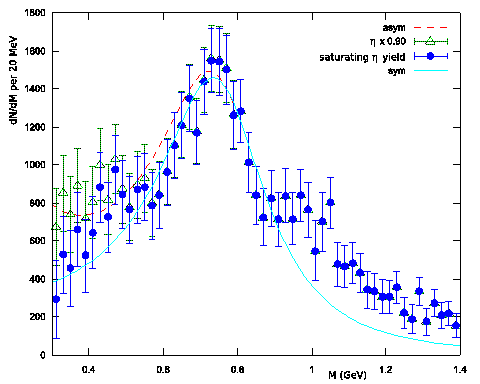}
 \caption{The spectrum of dileptons in semi-central collisions 
In(158 A GeV)-In measured by NA60-Collaboration \cite{damj} (points with error 
bars) compared to the $\rho $-meson peak in the medium with additional Cherenkov 
contribution (the dashed line). The solid line shows the Breit-Wigner shape 
with the modified width. Borrowed from \cite{drem}.}
\end{figure}

Several approaches have been advocated for explanation of the excess
\cite{pisa, hara, brow, bore, dusl, leup, rupp, elet, mart, rapp, hees}.
See the review \cite{haya}.
However, either positions, widths or heights presented some problems.
An alternative hypothesis about the role of Cherenkov gluon effect was promoted
in Ref. \cite{drem}. The necessary condition for Cherenkov effects to be 
observable within some energy interval is an excess of the refractivity index
of the medium $n$ over 1. According to general formulas this excess is 
proportional to the real part of the forward (depicted by 0 in formulas below) 
scattering amplitude. 
\begin{equation}
\Delta n ={\rm Re }n -1 \propto {\rm Re }F(M,0)>0. 
\label{delta}
\end{equation}
For the nuclear quark-gluon medium this requirement should be fulfilled for
the chromopermitivity of gluons \cite{drle}. The derivation of Eq. (\ref{bw}) 
shows that the real part is positive within the low-energy (left) wing of 
any resonance described by the Breit-Wigner formula. Therefore 
one could expect that the collective excitations of the quark-gluon medium may
contribute in these energy intervals in addition to the traditional
effects. Herefrom one gets the general prediction that the shape of
{\it any} resonance formed in nuclei collisions must become asymmetric with 
some excess within its left wing compared to the usual Breit-Wigner shape. 
Since the probability of Cherenkov radiation is proportional to $\Delta n$ 
the asymmetry must be proportional to it. Then the dilepton mass distribution 
must get the shape (the formula in \cite{drem} is slightly corrected):
\begin{equation}
\frac{dN_{ll}}{dM}=\frac {A}{(m_{r }^2-M^2)^2+m_r^2\Gamma ^2}
\left(1+w_r\frac{m_{r }^2-M^2}{m_r\Gamma }\Theta (m_{r}-M)\right)    \label{ll}
\end{equation}
The first term corresponds to the ordinary Breit-Wigner cross section depicted
by the solid line in Fig. 1 with a new width of 354 MeV obtained from the
fitting procedure. According to the optical theorem it is 
proportional to the imaginary part of the forward scattering amplitude. 
The second term is due to the coherent Cherenkov response of the medium 
proportional to the real part of the amplitude. It is in charge of the
observed asymmetry. It vanishes for $M>m_r$
because only positive $\Delta n$ lead to Cherenkov effects. Here, we take into 
account that the ratio of real to imaginary parts of Breit-Wigner amplitudes is
\begin{equation}
\frac {{\rm Re}F(M,0)}{{\rm Im}F(M,0)}=\frac {m_r^2-M^2}{m_r\Gamma }.
\label{ratio}
\end{equation}  
The weight of the second term to ordinary 
processes is described by the only adjustable parameter $w_r$ for a given 
resonance $r$. It must be found from comparison with experimental data. 
The sum is shown by the dashed line in Fig. 1. The fitted parameters are 
$A$=104 GeV$^3$, $\Gamma $=354 MeV, $w_{\rho }$=0.19. The shift of the 
$\rho $-meson mass is negligibly small. The width of the in-medium peak is about 
twice larger than that of the in-vacuum $\rho $-meson equal 150 MeV. 
The asymmetry parameter $w_{\rho }$ implies a reasonably small admixture 
of Cherenkov effects. It will be compared with the corresponding parameter $q$
in Fano treatment below.

Thus, Cherenkov effect asks for interaction of $\rho $-meson quarks with
some third component of a medium (gluon, collective modes...?). Such an 
interaction can result 
either in continuum or quasistable states. In quantum mechanics, the
interference of the continuum states and discrete levels of the reaction leads
to the well known Fano effect \cite{fano, fano1}. They interfere with
opposite phase on the two sides of the resonance as shown in \cite{fano1}.
The resonance asymmetry is described by the following formula \cite{fano1}:
\begin{equation}
\sigma = \frac {(q+\epsilon )^2}{1+\epsilon ^2}=1+\frac {q^2-1+2q\epsilon }
{\epsilon ^2 +1},
\label{sigma}
\end{equation}
where at the relativistic notation
\begin{equation}
\epsilon = \frac {M^2-m_r^2}{m_r\Gamma }.
\end{equation}
The parameter $q$ describes the relative strength of discrete states and
unperturbed continuum.
The asymmetric components are proportional to $M^2-m_r^2$ in both formulas
(\ref{ll}) and (\ref{sigma}). Equating them, one gets the relation between 
the parameters $q$ and $w_r$:
\begin{equation}
q=-w_{\rho }/2 \approx -0.1
\label{qw}
\end{equation}
which shows that the interference is not very strong but nevertheless quite
noticeable in the left wing (the negative sign!).

The admixture of the contribution of direct states to this effect compared
to the influence of the continuum was estimated in \cite{fano1} equal
$\pi q^2/2$ which amounts to about 0.014 in our case. Thus we conclude that
continuum scattering plays an overwhelming role while quasibound states are 
formed rather rarely.  

The interference of continuum and quasibound states is at the origin of
asymmetric resonances and of the phenomenologic prescription of $\Delta n>0$. 
The similar effect for the compound nuclei is known in nuclear physics as 
Feshbach resonances \cite{fesh}. In our case, the nature of the third component 
interacting with $\rho $-meson quarks and producing quasibound states
is however left unknown. With the above estimate of low admixture of the direct
states one is even tempted to speculate that electromagnetic forces could be
in charge of this effect. It is quite possible that such a component 
initiates the binding of two otherwise independent quarks in the
medium. The formation of the weakly bound triple-states in the collisions of 
three particles when two-particle forces are too weak to produce bound 
two-quark dimers is known as Efimov effect \cite{efim, ferl}. From the
theoretical side, models of (collective?) excitations in the nuclear
medium which help to get an insight to this problem are welcome.
From the experimental side, very little is still known about other resonances
but the low-mass asymmetry seems universal and gives some hope for further
progress. 

At the end, let us stress once again that the electromagnetic interaction of 
electrons with atoms in ordinary media with $\Delta n>0$ is described in quantum 
mechanics as interference of continuum and quasibound states. Thus, beside 
asymmetric resonances, it is in charge of famous Cherenkov rings of photons as 
well. The analogous explanation of {\it hadronic} ring-like events 
observed \cite{apan} in high energy nuclear collisions in terms of Cherenkov 
gluons was proposed in \cite{dr1, dr2}.

\medskip

{\bf Acknowledgments}

\medskip 
 
I am grateful for support by the RFBR-grant
14-02-00099  and the RAS-CERN program.


\begin{thebibliography}{99}
\bibitem{bw}
G. Breit, E. Wigner, Phys. Rev. {\bf 47}, 519 (1936).
\bibitem{rice}
O.K. Rice, J. Chem. Phys. {\bf 1}, 375 (1933).
\bibitem{fano}
U. Fano, Nuovo Cim. {\bf 12}, 156 (1935).
\bibitem{agak}
G. Agakichiev et al. (CERES) Phys. Rev. Lett. {\bf 75}, 1272 (1995);
Phys. Lett. B {\bf 422}, 405 (1998); Eur. Phys. J. C {\bf 41}, 475 (2005).
\bibitem{adam}
D. Adamova et al. (CERES) Phys. Rev. Lett. {\bf 91}, 042301 (2003);
{\bf 96}, 152301 (2006).
\bibitem{arna}
R. Arnaldi et al. (NA60) Phys. Rev. Lett. {\bf 96}, 162302 (2006).
\bibitem{damj}
S. Damjanovich et al. (NA60) Eur. Phys. J. C {\bf 49}, 235 (2007);
Nucl. Phys. A {\bf 783}, 327 (2007).
\bibitem{trnk}
D. Trnka et al. Phys. Rev. Lett. {\bf 94}, 192303 (2005).
\bibitem{naru}
M. Naruki et al. (KEK) Phys. Rev. Lett. {\bf 96}, 092301 (2006).
\bibitem{muto}
R. Muto at al. (KEK) Phys. Rev. Lett. {\bf 98}, 042501 (2007).
\bibitem{kozl}
A. Kozlov (PHENIX), nucl-ex/0611025.
\bibitem{kotu}
M. Kotulla (CBELSA/TAPS), nucl-ex/0609012.
\bibitem{tser}
I. Tserruya, Nucl. Phys. A {\bf 774}, 415 (2006).
\bibitem{drem}
I.M. Dremin, V.A. Nechitailo, Int. J. Mod. Phys. A {\bf 24}, 1221 (2009).
\bibitem{pisa}
R. Pisarski, Phys. Lett. B {\bf 110}, 155 (1982).
\bibitem{hara}
M. Harada, K. Yamawaki, Phys. Rep. {\bf 381}, 1 (2003).
\bibitem{brow}
G.E. Brown, M. Rho, Phys. Rev. Lett. {\bf 66}, 2720 (1991); Phys. Rep. {\bf 269}, 
333 (1996); Phys. Rep. {\bf 363}, 85 (2002).
\bibitem{bore}
K.G. Boreskov, J.H. Koch, L.A. Kondratyuk, M.I. Krivoruchenko, 
Yad. Fiz. {\bf 59}, 1908 (1996) [Phys. Atom. Nucl {\bf 59}, 1844 (1996)].
\bibitem{dusl}
K. Dusling, D. Teaney, I. Zahed, nucl-th/0604071.
\bibitem{leup}
S. Leupold, W. Peters, U. Mosel, Nucl. Phys. A {\bf 628}, 311 (1998).
\bibitem{rupp}
J. Ruppert, T. Renk, B. Muller, Phys. Rev. C {\bf 73}, 034907 (2006).
\bibitem{elet}
V.L. Eletsky, M. Belkacem, P.J. Ellis, J.L. Kapusta, Phys. Rev. C {\bf 64},
035202 (2001).
\bibitem{mart}
A.T. Martelli, J.P. Ellis, Phys. Rev. C {\bf 69}, 065206 (2004).
\bibitem{rapp}
R. Rapp, J. Wambach, Eur. Phys. J. A {\bf 6}, 415 (1999); Adv. Nucl. Phys.
{\bf 25}, 1 (2000).
\bibitem{hees}
H. van Hees, R. Rapp, Phys. Rev. Lett. {\bf 97}, 102301 (2006).
\bibitem{haya}
R.S. Hayano, T. Hatsuda, Rev. Mod. Phys. {\bf 82}, 2049 (2010).
\bibitem{drle}
I.M. Dremin, A.V. Leonidov, Physics-Uspekhi {\bf 53}, 1123 (2010).
\bibitem{fano1}
U. Fano, Phys. Rev. {\bf 124}, 1866 (1961).
\bibitem{fesh}
H. Feshbach, Ann. Phys. (N.Y.) {\bf 5}, 357 (1958).
\bibitem{efim}
V. Efimov, Phys. Lett. B {\bf 33}, 663 (1970).
\bibitem{ferl}
F. Ferlaino, R. Grimm, Physics {\bf 3}, 9 (2010).
\bibitem{apan}
A.V. Apanasenko et al.  JETP Lett. {\bf 30}, 145 (1979).
\bibitem{dr1}
I.M. Dremin, JETP Lett. {\bf 30}, 140 (1979).
\bibitem{dr2}
I.M. Dremin, Nucl. Phys. A {\bf 767}, 233 (2006).

\end{thebibliography}
\end{document}